\documentclass[trackchanges,twocolumn]{aastex701}

\newcommand{\twelveco}{$^{12}$CO}
\newcommand{\thirteenco}{$^{13}$CO}
\newcommand{\ceighteeno}{C$^{18}$O}
\usepackage{amsmath}
\usepackage{soul}
\usepackage{xcolor}
\definecolor{APPyellow}{HTML}{FFFDE7} 
\definecolor{APPorange}{HTML}{FFF3E0}
\definecolor{APPred}{HTML}{FFEBEE}
\definecolor{APPdeepred}{HTML}{F9EBEB}

\newcommand{\cn}{CN}

\begin{document}

\title{Turbulence Cascade in Cygnus~X Revealed by Multi-point VDF Method}

\author[orcid=0009-0006-5571-0191,gname=Junjie, sname=Huang]{Junjie Huang}
\affiliation{School of Astronomy and Space Science, Nanjing University, 163 Xianlin Avenue, Nanjing 210023, Jiangsu, People’s Republic of China}
\affiliation{Key Laboratory of Modern Astronomy and Astrophysics (Nanjing University), Ministry of Education, Nanjing 210023, Jiangsu, People’s Republic of China}
\email{221240085@smail.nju.edu.cn} 

\author[orcid=0009-0008-9415-3329,gname=Yangjun, sname=Pu]{Yangjun Pu} 
\affiliation{School of Astronomy and Space Science, Nanjing University, 163 Xianlin Avenue, Nanjing 210023, Jiangsu, People’s Republic of China}
\affiliation{Key Laboratory of Modern Astronomy and Astrophysics (Nanjing University), Ministry of Education, Nanjing 210023, Jiangsu, People’s Republic of China}
\email{221840318@smail.nju.edu.cn}

\author[orcid=0000-0002-5093-5088,gname=Keping,sname=Qiu]{Keping Qiu$^*$}
\affiliation{School of Astronomy and Space Science, Nanjing University, 163 Xianlin Avenue, Nanjing 210023, Jiangsu, People’s Republic of China}
\affiliation{Key Laboratory of Modern Astronomy and Astrophysics (Nanjing University), Ministry of Education, Nanjing 210023, Jiangsu, People’s Republic of China}
\email{kpqiu@nju.edu.cn}

\author[orcid=0000-0002-4774-2998,gname=Junhao, sname=Liu]{Junhao Liu}
\affiliation{School of Astronomy and Space Science, Nanjing University, 163 Xianlin Avenue, Nanjing 210023, Jiangsu, People’s Republic of China}
\affiliation{Key Laboratory of Modern Astronomy and Astrophysics (Nanjing University), Ministry of Education, Nanjing 210023, Jiangsu, People’s Republic of China}
\email{liujunhao@nju.edu.cn}

\author[orcid=0009-0008-3635-1167,gname=Yingxi, sname=Li]{Yingxi Li}
\affiliation{School of Astronomy and Space Science, Nanjing University, 163 Xianlin Avenue, Nanjing 210023, Jiangsu, People’s Republic of China}
\affiliation{Key Laboratory of Modern Astronomy and Astrophysics (Nanjing University), Ministry of Education, Nanjing 210023, Jiangsu, People’s Republic of China}
\email{starryyousei@gmail.com}

\author[orcid=0000-0003-0596-6608,gname=Mengke, sname=Zhao]{Mengke Zhao}
\affiliation{School of Astronomy and Space Science, Nanjing University, 163 Xianlin Avenue, Nanjing 210023, Jiangsu, People’s Republic of China}
\affiliation{Key Laboratory of Modern Astronomy and Astrophysics (Nanjing University), Ministry of Education, Nanjing 210023, Jiangsu, People’s Republic of China}
\email{mkzhao@nju.edu.cn}

\correspondingauthor{Keping Qiu}
\email{kpqiu@nju.edu.cn}

\begin{abstract}

Turbulence plays a crucial role in regulating star formation activities within molecular clouds, yet few methods can directly reveal its properties and underlying processes. We use molecular line data from the Nobeyama 45m Cygnus~X CO Survey to study the turbulence properties and their relationships with star-forming activities and/or other non-thermal motions. In this work, we apply the multi-point velocity dispersion function (VDF), rather than direct linewidth measurements, to investigate the non-thermal properties of molecular cloud motions. We filter out the large-scale ordered structure and isolate a relatively small-scale turbulence component. Through the Friends In Velocity (FIVe) algorithm, we identify 10 substructures of the clouds and derive the turbulent properties of each cloud using the VDF method. We find that both the cloud-complex regions and the 10 velocity substructures exhibit turbulence correlation lengths of $\sim 2$--5 pc. This plateau scale suggests a parsec-scale turbulence correlation or driving scale in Cygnus~X. Below this scale, the rising VDFs trace the velocity scaling of the turbulent cascade, whereas larger-scale VDF variations likely reflect cloud-scale motions. The comparison between cloud complexes and substructures further suggests that, in observational data, the VDF may constrain the turbulence correlation scale more robustly than the turbulence velocity dispersion.

\end{abstract}


\section{Introduction} \label{sec:introduction}

Star formation is influenced by multiple physical factors, including self-gravity, thermal pressure, magnetic fields, turbulence, and feedback \citep[e.g.][]{2007ARA&A..45..481Z,2007ARA&A..45..565M,2014PhR...539...49K,2023ASPC..534..193P}.
Among these factors, turbulence can resist gravitational collapse, thereby reducing the rates of star formation. Additionally, it regulates the hierarchical fragmentation process from large scales to small scales \citep{2004RvMP...76..125M}. Furthermore, turbulence can induce strong gas compression through shocks in dense regions of the interstellar medium (ISM), where potential star formation may occur \citep{2024JApA...45...17S}. Measuring the physical properties of turbulence through spectral line observations is crucial for understanding the role of turbulence in molecular clouds.

In the past, the properties of turbulence in molecular clouds were extensively investigated using the linewidth--size relation, first identified observationally by \citet{1981MNRAS.194..809L}. He found that the velocity dispersion scales with cloud size approximately as
\begin{equation}
    \sigma \propto L^{0.38}.
\end{equation}
Subsequent observations have reported linewidth--size exponents ranging from $\sim 0.2$ to $0.7$ \citep[e.g.][]{1983ApJ...270..105M,1987ApJ...319..730S,1995ApJ...446..665C,1998ApJ...504..223G,2004ApJ...615L..45H}, while related Larson-type scaling relations have also been discussed in terms of the universality of molecular cloud structure \citep{2010A&A...519L...7L}. These empirical scalings are often interpreted in connection with turbulent cascades, but the theoretical picture is not unique. The classical Kolmogorov model predicts a velocity scaling exponent of $1/3$ for isotropic, incompressible, non-magnetic turbulence, whereas molecular clouds are highly compressible, magnetized, and usually supersonic. In shock-dominated or Burgers-like turbulence, the expected exponent is closer to $1/2$ \citep{2004ARA&A..42..275S,2007ARA&A..45..565M}. Numerical simulations of supersonic, approximately isothermal turbulence further show that the observed scaling and density structure depend on the Mach number, magnetic field, driving scale, and driving mode \citep{1997ApJ...474..292V,2002ApJ...570..734B,2010A&A...512A..81F}. In addition, thermally bistable atomic flows may already imprint cloud-like structures before or during the formation of molecular gas \citep{2007A&A...465..431H,2007A&A...465..445H}.

Although the linewidth--size relation provides a useful empirical diagnostic of turbulent motions, its interpretation can be complicated in massive molecular clouds, as the velocity dispersion in massive clumps significantly deviates from the Larson relation \citep{1997ApJ...476..730P,2003ApJS..149..375S}. 
Moreover, \citet{2009ApJ...699.1092H} found a positive correlation between linewidth and column density in giant molecular clouds, indicating that internal motions increase with surface density and are consistent with virial support, rather than a universal turbulent cascade. In addition, observed linewidths can be affected by non-turbulent kinematic components, such as rotation, inflow, expansion, and outflow. These effects can bias the use of linewidths as direct tracers of turbulence, motivating alternative approaches that characterize the velocity field across spatial scales.

The velocity centroid dispersion function, or structure function, can be used to investigate the properties of turbulence in molecular clouds across different spatial scales \citep{1985ApJ...295..479D,1994ApJ...429..645M}. 
These functions are insensitive to thermal broadening \citep{2022ApJ...935...77L} and are not affected by line-of-sight integration \citep{2016ApJ...821...21C}. The velocity field comprises large-scale and small-scale components, each of which may consist of both ordered and turbulent motions. \cite{2019ApJ...874...75C} proposed that multi-point structure functions can effectively remove the large-scale ordered component and verified it using numerical simulation data. \cite{2023ApJ...949...30L} have successfully applied this method to astronomical observation data and extracted supersonic turbulence on scales of $\sim 0.01$ pc.

In this work, we apply the multi-point velocity dispersion functions (VDFs) to Cygnus~X to investigate the turbulent properties and structures of the cloud. This paper is organized as follows. In Section 2, we describe the data used in this study. In Section 3, we introduce the VDF method, describe the FIVe algorithm used to identify velocity substructures, and present the VDF results for both the overall cloud and individual substructures. In Section 4, we discuss the implications of the VDFs for the turbulent cascade. Finally, in Section 5, we summarize our main results.

\section{Data} \label{sec:data}

Cygnus~X is one of the largest and most active high-mass star-forming regions in the Milky Way \citep[see the review by][]{2008hsf1.book...36R}, hosting numerous H\,II regions \citep{1991A&A...241..551W}, OB associations \citep{2001A&A...371..675U}, and high-mass star-forming cores \citep{2002A&A...384..225S,2004ApJ...601..952S,2007A&A...476.1243M,2019ApJS..241....1C}. The distance and physical association of the molecular clouds in Cygnus~X have been debated for decades. Based on large-scale CO observations, \citet{2006A&A...458..855S} suggested that Cygnus~X is a physically connected molecular complex. This picture was later confirmed by maser-parallax measurements, which placed most maser sites in Cygnus~X at distances of $\sim 1.3$--1.6~kpc, with Cygnus~X North located at approximately 1.4~kpc \citep{2012A&A...539A..79R,2013ApJ...769...15X}. The goal of our study is to investigate the turbulence characteristics of the molecular cloud as a whole; thus, the local distance uncertainties within the cloud are not expected to significantly affect our results. For simplicity, we adopt a uniform distance of 1.4~kpc.

The data analyzed in this study were obtained from the Nobeyama 45m Cygnus~X CO Survey\footnote[1]{\url{https://www.nro.nao.ac.jp/~nro45mrt/html/results/data.html}} \citep{2018ApJS..235....9Y, 2019ApJ...883..156T}. We use observations of \twelveco($J=1-0$), \thirteenco($J=1-0$), \ceighteeno($J=1-0$), and \cn($N=1-0$), covering both Cygnus~X North and South. The \twelveco($J=1-0$) line traces relatively diffuse molecular gas over a large volume, but is often optically thick and susceptible to self-absorption and line-of-sight contamination. Although its optically thin critical density is $\sim10^3~{\rm cm^{-3}}$, radiative trapping allows emission from gas below $10^2~{\rm cm^{-3}}$ \citep{2015PASP..127..299S}. The \thirteenco($J=1-0$) line traces gas at a similar characteristic density but is generally less optically thick, providing clearer velocity components across the cloud. The weaker \ceighteeno($J=1-0$) emission offers insufficient spatial coverage, while \cn($N=1-0$), with a critical density of order $10^5~{\rm cm^{-3}}$, mainly traces dense or UV-irradiated gas \citep{2015PASP..127..299S}. We therefore choose the \thirteenco($J=1-0$) data as the tracer for the VDF analysis.

In this work, we use the publicly available data cube smoothed to an effective angular resolution of $46^{\prime\prime}$, following \citet{2018ApJS..235....9Y}. This smoothing improves the signal-to-noise ratio (S/N) and helps mitigate low-level striping artifacts. The adopted cube has a velocity-channel width of $1.0~\mathrm{km\,s^{-1}}$ and a median rms noise of 0.20~K in $T_{\mathrm{mb}}$. Although this spectral resolution limits our sensitivity to sub-km~s$^{-1}$ kinematic structures, our analysis focuses on the statistical properties of cloud-scale velocity fluctuations rather than on resolving individual narrow velocity components. The effects of angular and velocity resolution on the derived VDFs are assessed in the Appendix.

\section{Results} \label{sec:results}

In this section, we present our methodology and its application to the Cygnus~X molecular cloud. We then characterize the turbulent properties across different spatial scales and regions, and explore their connection to star-forming activity.

\subsection{Moment 0 and Velocity Centroid} \label{sec:m0a1}

To explore the kinematics of Cygnus~X, we first calculate the moment~0 and velocity centroid at each position $\boldsymbol{x}$ using the \thirteenco\ data from the Nobeyama 45m Cygnus~X CO Survey.

The traditional method for calculating moment 0 involves summing the intensity of all velocity channels, as follows:
\begin{equation}
    M_{0}(\boldsymbol{x}) = \sum_{i}^{N_{\mathrm{ch}}} I_{i}(\boldsymbol{x}) \Delta v_{\mathrm{ch}}
\end{equation}
Similarly, the velocity centroid (i.e., moment~1) is calculated as:
\begin{equation}
    V_{c}(\boldsymbol{x}) = \dfrac{\sum_{i}^{N_{\mathrm{ch}}} I_{i}(\boldsymbol{x}) v_{i} \Delta v_{\mathrm{ch}}}{\sum_{i}^{N_{\mathrm{ch}}} I_{i}(\boldsymbol{x}) \Delta v_{\mathrm{ch}}}
\end{equation}
Where $I_{i}(\boldsymbol{x})$, $v_{i}$, $\Delta v_{\mathrm{ch}}$, and $N_{\mathrm{ch}}$ represent the line intensity, line-of-sight velocity, channel width, and number of integrated channels, respectively. However, this method cannot effectively suppress the impact of noise from individual velocity channels. We therefore select a small patch of sky background as a reference region for noise estimation. For each velocity channel, we then compute the noise (i.e., the root mean square, rms) and identify channels with intensities greater than $10\sigma$ as valid channels. The velocity centroid for each pixel is computed using only these valid channels.

\begin{figure*}[htbp]
    \centering
    \includegraphics[width=0.9\textwidth]{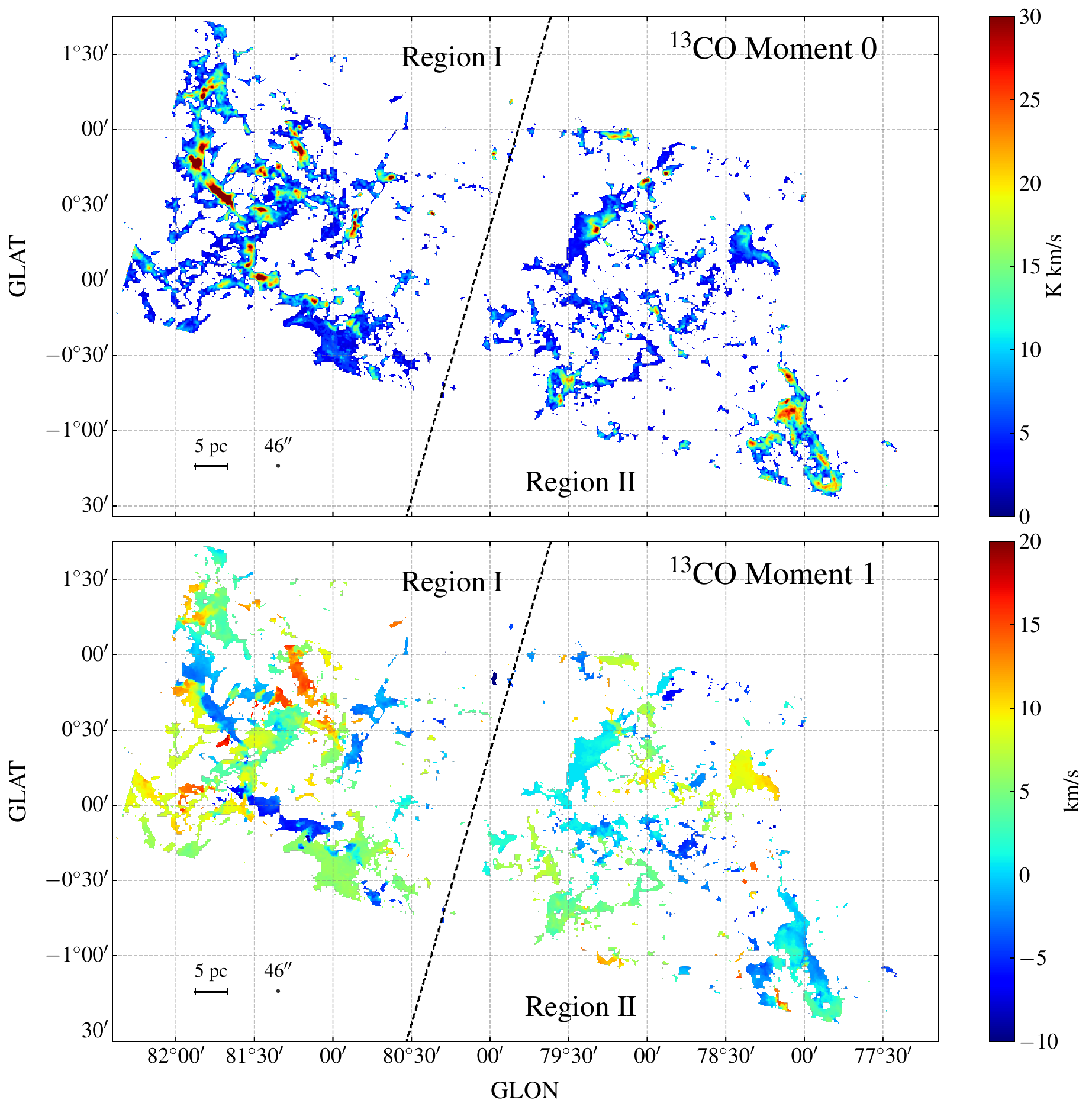}
    \caption{Moment 0 (upper panel) and velocity centroid (lower panel) maps of Nobeyama \thirteenco($J=1-0$) line emission toward the Cygnus~X complex. The dashed line separates Region~I (Cygnus~X North) and Region~II (Cygnus~X South). The scale bar and the effective angular resolution are indicated in the lower left corners of each panel.}
    \label{fig:m}
\end{figure*}

In practice, we only consider the line emission from $-$10 to 20~$\mathrm{km\,s^{-1}}$, which encompasses the main velocity component of Cygnus~X, as indicated by the previous works \citep{2006A&A...458..855S,2016A&A...587A..74S}. Fig.~\ref{fig:m} shows the maps of moment 0 and velocity centroid. Both maps reveal a clear gap between the two main molecular complexes. This gap is located near the Cyg~OB2 association and may be related to the clearing of molecular gas by massive-star feedback. We therefore divide Cygnus~X into two regions for further analysis: Region~I, corresponding approximately to Cygnus~X North, and Region~II, corresponding approximately to Cygnus~X South.

\subsection{Multi-point Velocity Centroid Dispersion Functions} \label{sec:vdf}

Following \cite{2023ApJ...949...30L}, we calculate the multi-point structure functions for the velocity centroid. The resulting 2-point to 5-point VDFs are as follows:

\begin{subequations}
\begin{align}
&\begin{aligned}
\operatorname{VDF}_{2\mathrm{pt}}(l) &= \Big\langle \Big| V_{c}(\boldsymbol{x}+\boldsymbol{l}) - V_{c}(\boldsymbol{x}) \Big|^{2} \Big\rangle^{1/2}
\end{aligned}
\\
&\begin{aligned}
\operatorname{VDF}_{3\mathrm{pt}}(l) &= \frac{1}{\sqrt{3}} \Big\langle \Big| V_{c}(\boldsymbol{x}-\boldsymbol{l}) - 2 V_{c}(\boldsymbol{x}) \\
&+ V_{c}(\boldsymbol{x}+\boldsymbol{l}) \Big|^{2} \Big\rangle^{1/2}
\end{aligned}
\\
&\begin{aligned}
\operatorname{VDF}_{4\mathrm{pt}}(l) &= \frac{1}{\sqrt{10}} \Big\langle \Big| V_{c}(\boldsymbol{x}-\boldsymbol{l}) - 3 V_{c}(\boldsymbol{x})  \\
&+ 3 V_{c}(\boldsymbol{x}+\boldsymbol{l}) - V_{c}(\boldsymbol{x}+2\boldsymbol{l}) \Big|^{2} \Big\rangle^{1/2}
\end{aligned}
\\
&\begin{aligned}
\operatorname{VDF}_{5\mathrm{pt}}(l) &= \frac{1}{\sqrt{35}} \Big\langle \Big| V_{c}(\boldsymbol{x}-2\boldsymbol{l}) - 4 V_{c}(\boldsymbol{x}-\boldsymbol{l}) \\
&+ 6 V_{c}(\boldsymbol{x}) - 4 V_{c}(\boldsymbol{x}+\boldsymbol{l}) + V_{c}(\boldsymbol{x}+2\boldsymbol{l}) \Big|^{2} \Big\rangle^{1/2}
\end{aligned}
\end{align}
\end{subequations}

Here, $l$ denotes the separation interval. To suppress beam-smoothing effects on small scales \citep{1994ApJ...429..645M}, we define the minimum separation as $l_{\mathrm{res}}$ and increase $l$ in steps of $l_{\mathrm{res}}/2$, approximately following Nyquist sampling. For each VDF at scale $l$, all point pairs with separations in the range $[l-l_{\mathrm{res}}/4,\, l+l_{\mathrm{res}}/4]$ are assigned to that bin. To avoid overly broad intervals and sparse sampling at large separations, we require each maximum-$l$ bin to contain more than 1000 point pairs. In our data, $l_{\mathrm{res}} \approx 0.312$~pc, corresponding to a step size of 0.156~pc. Because each distance bin contains a sufficiently large number of point pairs, the statistical uncertainties in the VDFs are negligible \citep{1994ApJ...429..645M}; therefore, error bars are omitted from the VDF plots.

Theoretically, an $n$-point VDF can eliminate large-scale structures up to order $n-2$ (see \citealt{2019ApJ...874...75C}). If the VDF effectively removes the large-scale structure, its profile is expected to behave as follows: first, due to the turbulent cascade process---where energy is transferred from large to small scales and ultimately dissipated---the VDF increases following a power-law relation. Once the separation exceeds the turbulence correlation scale $l_{s}$, the VDFs flatten, forming a ``plateau''. Here, $l_s$ denotes the characteristic scale beyond which the velocity correlation function, $R(l)=\langle v_s(x)v_s(x+l)\rangle$, becomes negligible. It can therefore be interpreted as the largest correlated turbulent scale sampled by the observations. If energy injected on larger scales is transferred into turbulent motions, the corresponding driving scale may be comparable to $l_s$ \citep{pope2001turbulent}. The height of the plateau is expected to be $\sqrt{2} \sigma_{s}$, where $\sigma_{s}$ is the rms of the turbulence velocity field. As the separation increases further, the VDF may rise again and even exhibit fluctuations, indicating that larger-scale higher-order velocity structures, which cannot be fully removed by a finite-order multi-point VDF, become increasingly important.

The simulations presented in \cite{2019ApJ...874...75C} include large-scale ordered motions together with small-scale turbulent fluctuations, providing a controlled and idealized framework for examining the behavior of the VDF, in which multi-point VDFs effectively remove large-scale ordered components. In observational data, however, the kinematics are more complex, potentially containing both small-scale ordered motions and large-scale turbulent motions. For lower-point VDFs, contributions from small-scale ordered motions may persist at scales below the correlation length $l_{s}$, leading to an increase in the VDF. In contrast, higher-point VDFs are expected to be more effective at removing small-scale ordered motions as well, such that the resulting plateau robustly reflects the intrinsic small-scale turbulent component. On the other hand, the role of large-scale turbulent motions in shaping the VDF remains poorly constrained. As a result, the increase in the VDF beyond the plateau cannot be uniquely attributed to large-scale ordered motions, and the coupling between large- and small-scale turbulence remains unclear. A broader and more complete kinematic range of observations is therefore needed to clarify how large-scale turbulence influences the VDF.

The multi-point VDF can also be used to separate the large-scale ($V_{c,l}$) and small-scale ($V_{c,s}$) structures in the velocity centroid map \citep{2019ApJ...874...75C}. For example, if VDF$_{\mathrm{5pt}}$ shows a plateau near $r_{p}$, the averaged approximation for the large-scale component $V_{c,l}$ is:
\begin{equation}
\begin{aligned}
    \langle V_{c,l}(\boldsymbol{x}) \rangle &= \sum_{r_{p}-\Delta < |\boldsymbol{r}| < r_{p}+\Delta} \Big[ 4V_{c}(\boldsymbol{x}+\boldsymbol{r}) + 4V_{c}(\boldsymbol{x}-\boldsymbol{r}) \\
    &- V_{c}(\boldsymbol{x}+2\boldsymbol{r}) - V_{c}(\boldsymbol{x}-2\boldsymbol{r}) \Big] / 6N
\end{aligned}
\end{equation}
Here, $r_{p}-\Delta$ and $r_{p}+\Delta$ fall within the plateau range, and $N$ denotes the total number of points. Consequently, the turbulence velocity field is estimated as:
\begin{equation}
    V_{c,s}(\boldsymbol{x}) \simeq V_{c}(\boldsymbol{x}) - \langle V_{c,l}(\boldsymbol{x}) \rangle
\end{equation}
Using the derived turbulence velocity field, we can calculate $\sigma_{s}$ and compare the result with the plateau of the VDFs.

\subsection{Turbulence in Region~I and II} \label{sec:turb-large}

\begin{figure*}[htbp]
    \centering
    \includegraphics[width=0.9\textwidth]{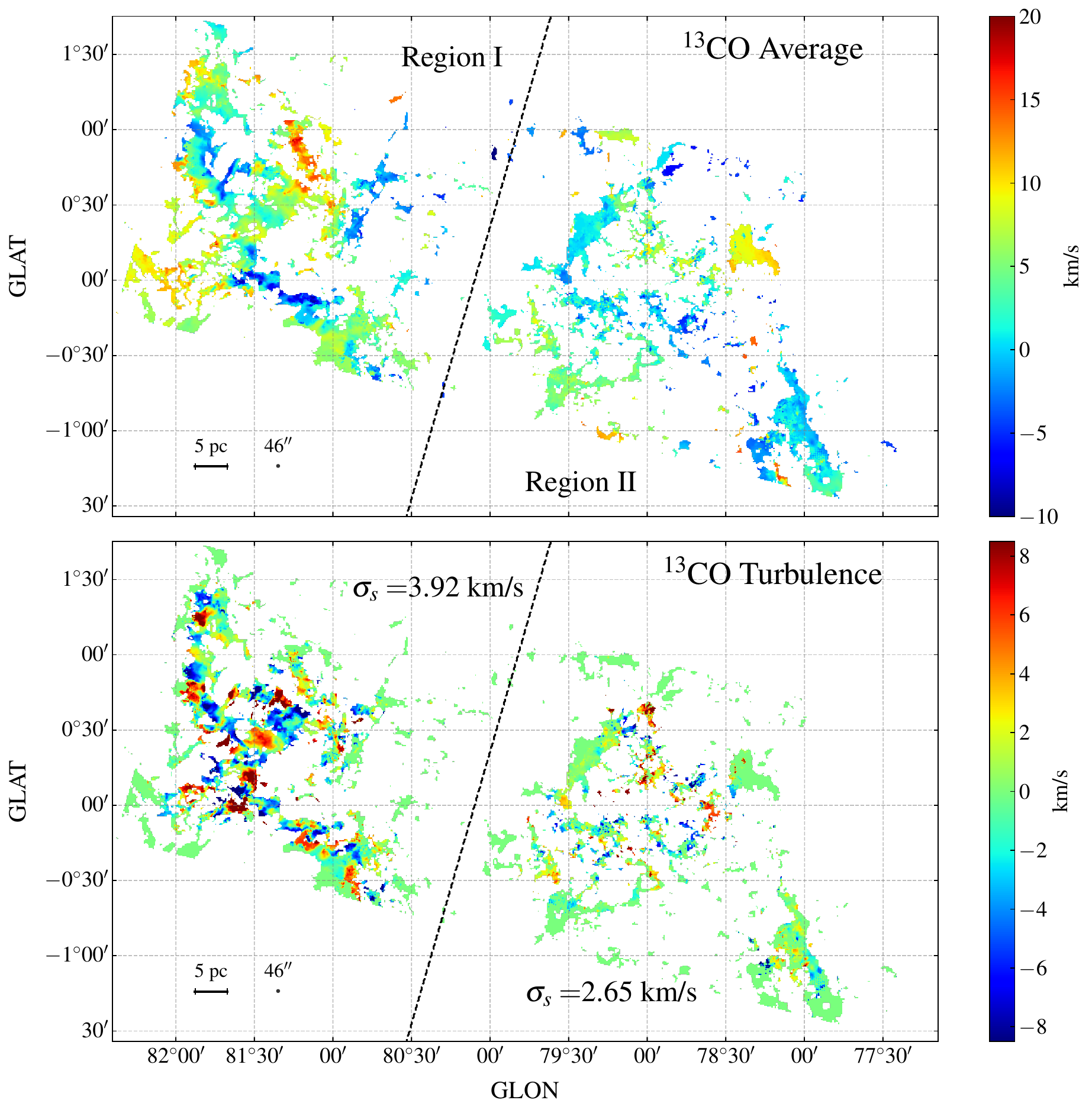}
    \caption{The large-scale averaged velocity centroid map (upper panel) and residual small-scale turbulence velocity centroid map (lower panel) revealed by Nobeyama \thirteenco\ $(J=1-0)$ line emission toward the Cygnus~X complex. The dashed line separates Region~I (Cygnus~X North) and Region~II (Cygnus~X South). In the lower panel, we also calculate and label the standard deviation of the turbulence component. The scale bar and the effective angular resolution are indicated in the lower left corners of each panel.}
    \label{fig:turb}
\end{figure*}

After obtaining the VDFs from the Nobeyama \thirteenco\ line data, we try to separate the large-scale and small-scale structures. We apply the method in Sec.~\ref{sec:vdf} to the entire Cygnus~X Region~I and Region~II. Specifically, for each valid pixel in the velocity centroid map (Fig.~\ref{fig:m}), we use Eq.~(5) to compute the averaged large-scale component $V_{c,l}$. From a mathematical point of view, Eq.~(5) performs a weighted average over two annuli centered on each pixel, with radii $r_{p}$ and $2r_{p}$ and widths $2\Delta$, which acts as a dual-ring-like filter extracting the large-scale velocity component. Then with Eq.~(6), we could finally get the estimated small-scale velocity field. The results are shown in Fig.~\ref{fig:turb}. The upper panel displays the large-scale averaged velocity centroid map, and the lower panel shows the residual small-scale velocity centroid map. As discussed above, when using 4- or 5-point VDFs, the ordered velocity components, including those on small scales, can be largely suppressed. Therefore, the lower panel of Fig.~\ref{fig:turb} can be regarded as an approximate estimate of the small-scale turbulent velocity component. On larger spatial scales, however, the velocity variations are again dominated by large-scale components, regardless of whether they are ordered or turbulent.

For the turbulent component, we also compute the standard deviation in the two regions, obtaining $\sigma_{s}=3.92~\mathrm{km\,s^{-1}}$ for Region~I and $\sigma_{s}=2.65~\mathrm{km\,s^{-1}}$ for Region~II, as labeled in the lower panel of Fig.~\ref{fig:turb}. Region~I therefore shows a broader range and larger amplitude of residual velocity fluctuations than Region~II. This difference is also visible in the map, where paired red-shifted and blue-shifted residual structures are more prominent in Region~I.

Region~I corresponds approximately to Cygnus~X North and contains the DR21/W75N region, one of the most active high-mass star-forming environments in Cygnus~X. Previous studies have identified numerous massive dense cores \citep{2007A&A...476.1243M}, powerful molecular outflows \citep{2013A&A...558A.125D,2025A&A...697A.186K}, and a network of filaments that merge with and feed the DR21 ridge \citep{2010A&A...520A..49S,2012A&A...543L...3H,2021ApJ...908...70H,2022ApJ...927..106C,2026AJ....171..145Y}. Recent [C~{\sc ii}], H~{\sc i}, and CO observations further suggest that molecular and atomic gas components in the DR21/W75N region are dynamically interacting over a broad velocity range \citep{2023NatAs...7..546S}. By comparison, Region~II, corresponding approximately to Cygnus~X South, is more diffuse and contains fewer prominent high-mass star-forming structures. The two regions therefore provide contrasting physical environments for comparing their residual velocity fields and VDF properties.

\begin{figure*}[htbp]
    \centering
    \includegraphics[width=0.8\textwidth]{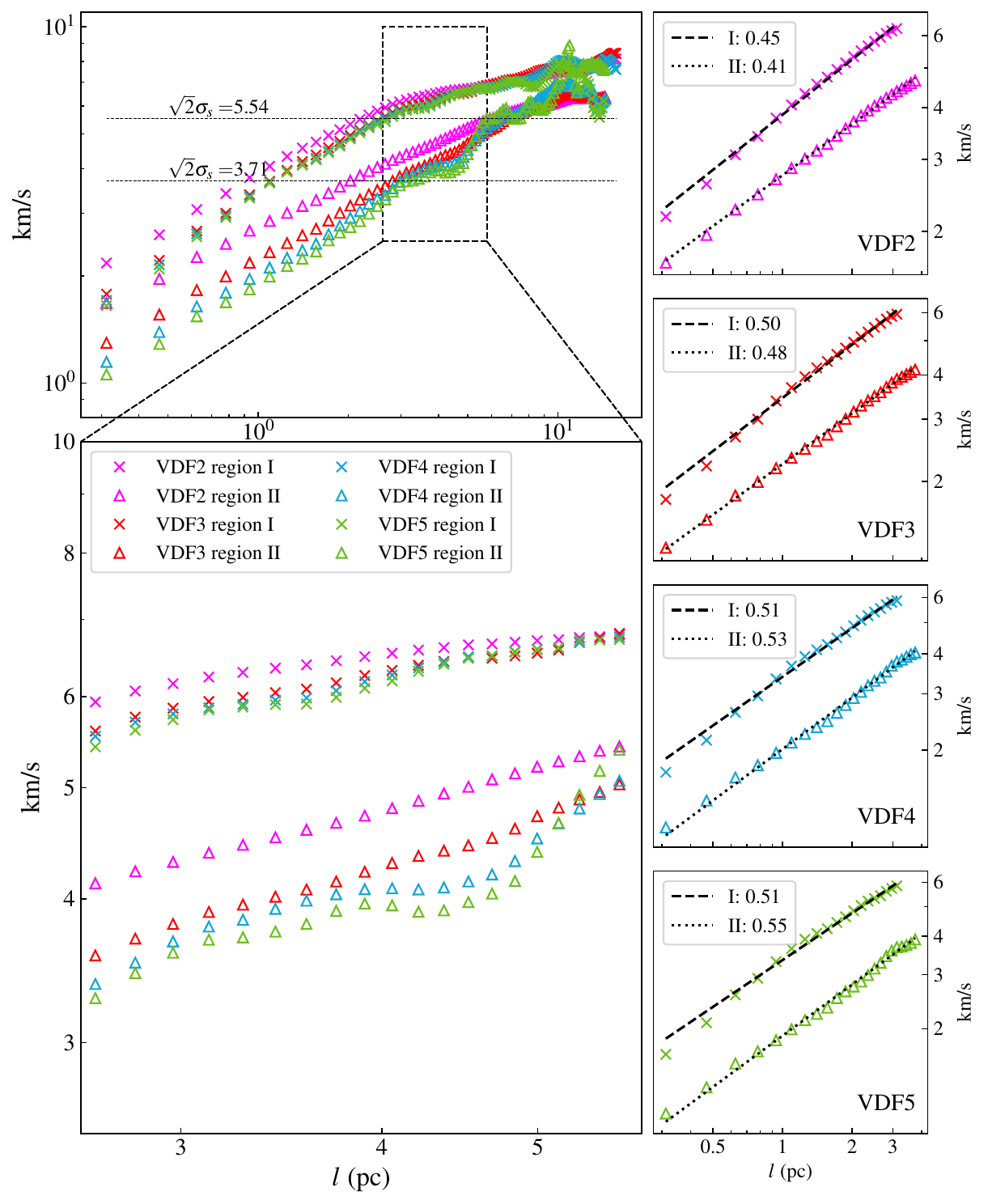}
    \caption{The 2- to 5-point velocity dispersion functions (VDFs) for Region~I and Region~II from \thirteenco\ line emission in Cygnus~X. The upper left panel shows the entire VDF plot, with crosses denoting Region~I and triangles denoting Region~II. Different colors represent different VDFs: pink for 2-point, red for 3-point, blue for 4-point, and green for 5-point VDFs. The horizontal dashed lines represent $\sqrt{2}$ times $\sigma_{s}$ of turbulence calculated in Fig.~\ref{fig:turb}. The zoomed-in region for the plateau is enclosed by the dashed rectangle and presented in the lower left panel. The right four sub-plots show 2- to 5-point VDFs before the plateau scale, with power-law fits (black dashed lines for Region~I and dotted lines for Region~II), and the power exponent labeled.}
    \label{fig:vdfl}
\end{figure*}

Fig.~\ref{fig:vdfl} shows the results of the multi-point VDFs. The entire VDF profiles and a zoomed-in view of the plateau ranges are shown in the left panels, while the right panels display the VDFs at separations below the plateau together with power-law fits. The horizontal dashed lines indicate $\sqrt{2}\sigma_s$, where $\sigma_s$ is estimated from the residual velocity fields shown in Fig.~\ref{fig:turb}.

The observational results are clear, with plateaus identified in both regions. In Region~I, the plateau extends over scales of $\sim 3$--4 pc, whereas in Region~II it spans from $\sim 4$--5 pc. The heights of these plateaus are consistent with the predictions from the turbulence map (the horizontal dashed line in the upper-left panel of Fig.~\ref{fig:vdfl}). We find that only the 4- and 5-point VDFs can reliably identify the plateau in our analysis. In contrast, a high-resolution ALMA study of NGC~6334~IV using the OCS (19--18) and $^{13}$CS (5--4) lines showed that even the 3-point VDF is sufficient to reveal a plateau on much smaller scales, $\sim 0.01$--0.03 pc \citep{2023ApJ...949...30L}. This difference may reflect the distinct physical properties and kinematic complexity of the two systems. Cygnus~X encompasses a much larger and more complex cloud system containing multiple velocity components and large-scale coherent structures, whereas NGC~6334~IV is a compact star-forming clump whose physical size is about two orders of magnitude smaller than that of the Cygnus~X region studied here.

The VDFs in both regions exhibit the expected behavior described in Section~\ref{sec:vdf}: they first increase approximately as a power law, reach a plateau at scales of $\sim 3$--5~pc, and subsequently rise again and fluctuate at larger separations. The VDF amplitudes in Region~I are systematically higher than those in Region~II, possibly reflecting its more active star-forming and dynamical environment. The few-parsec plateau is broadly consistent with energy injection by processes operating on comparable scales, such as the expansion of H~II regions and accretion or interactions within parsec-scale filamentary networks \citep{2006ApJ...653..361K,2012MNRAS.427..625W,2023NatAs...7..546S,2026A&A...707A.194L}. By contrast, the renewed rise of the VDFs beyond the plateau may include contributions from larger-scale velocity structures associated with cloud-assembly or converging flows, supernova-driven motions, and galactic shear or rotation \citep{2010A&A...520A..17K,2016ApJ...822...11P,2018ApJ...855...81S,2009ApJ...700..358T,2011ApJ...730...11T}. These associations are based primarily on the characteristic spatial scales of the different processes, and it is not applicable to identify the dominant driving mechanism solely based on the VDF analysis.

\subsection{Substructures within Cygnus~X}

The moment~1 map shows a highly structured and discontinuous velocity-centroid field, and many spectra contain multiple velocity peaks, indicating that several kinematic components overlap along the line of sight. This complexity is consistent with previous studies showing that the Cygnus~X complex might contain multiple gas layers, including foreground Cygnus Rift material and molecular structures at distances of approximately $1.3$--$1.6~\mathrm{kpc}$ \citep{2006A&A...458..855S,2007A&A...474..873S,2012A&A...541A..79G,2024AJ....167..220Z}. A VDF analysis of Regions~I and II as a whole may therefore mix distinct velocity components. We consequently use velocity-resolved spectra to identify kinematically coherent substructures and analyze their VDF properties separately.

The problem of separating velocity structures in molecular clouds is in some respects analogous to the identification of high-redshift galaxy clusters. Building on the simple Friends-of-Friends (FOF) algorithm originally developed in astrophysics \citep{1982ApJ...257..423H}, \citet{2013A&A...554A..55H} proposed the Friends In Velocity (FIVe) algorithm to identify multiple components within molecular clouds. Before identifying substructures, we extracted the spectrum at each spatial position and performed Gaussian fits to the line profiles. Depending on the peak intensity, up to three velocity components were identified at each position. These components were then used as input for the subsequent FOF analysis. The algorithm consists of the following four main steps:

\begin{enumerate}
    \item We first identify seed points in the position--position--velocity (PPV) cube using the following criteria: S/N $\geq 15$; at least 4 of the 8 surrounding points have S/N $\geq 10$; and a velocity gradient smaller than $1.8~\mathrm{km\,s^{-1}\,pc^{-1}}$, corresponding to a velocity variation on the order of the sound speed ($0.26~\mathrm{km\,s^{-1}}$) over the adopted spatial scale.
    
    \item For each seed, we perform an FOF search within a box of $60^{\prime\prime}$, using the same velocity-gradient criterion.
    
    \item We then assign isolated points, or those with S/N $< 15$, to the cloud components identified in the previous step. For each assignment, we require that the velocity gradient satisfy the same threshold of $1.8~\mathrm{km\,s^{-1}\,pc^{-1}}$.
    
    \item Finally, we discard components containing fewer than 1200 points (1.5\% of the total sample) to ensure sufficient statistical sampling for stable VDF estimates.
\end{enumerate}

\begin{figure*}[htbp]
    \centering
    \includegraphics[width=0.9\textwidth]{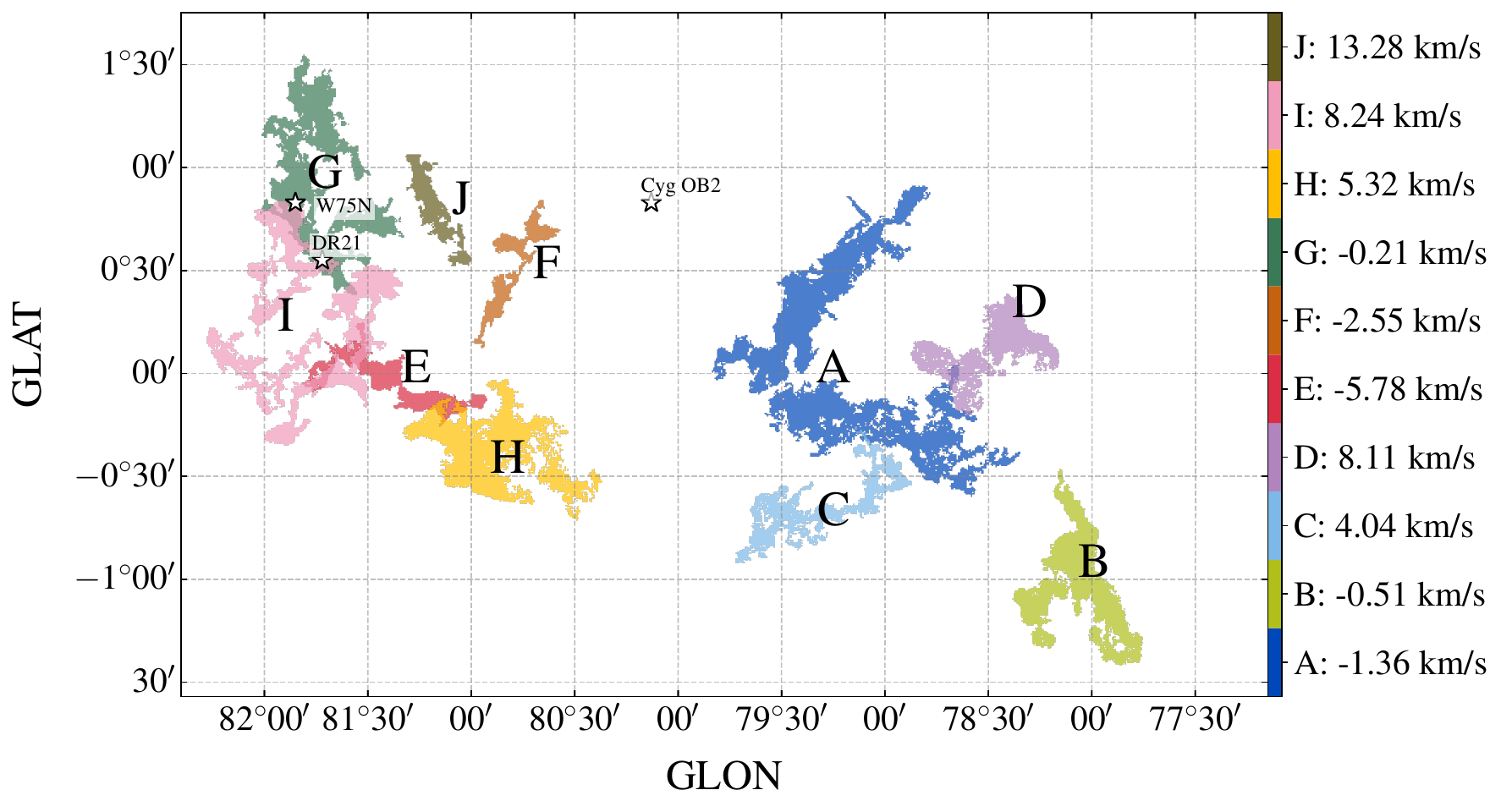}
    \caption{Spatial distribution and average velocity in each velocity substructure in Cygnus~X. The position and range of each component are marked with distinct colors. The right panel displays the average velocity for the ten substructures. The approximate projected locations of the DR21 ridge, the W75N complex, and the Cyg~OB2 association are also indicated for reference.}
    \label{fig:struct}
\end{figure*}

Using the FIVe algorithm, we obtained a map of the velocity substructures in Cygnus~X. As shown in Fig.~\ref{fig:struct}, we identify ten major velocity-coherent components with approximately compact or filamentary projected morphologies, six of which are located in Region~I and four in Region~II. The two regions are separated by a molecular-gas cavity wider than $30^{\prime}$, within which the Cyg~OB2 association is located. These substructures exhibit substantial differences in velocity regardless of their positions or morphologies, with mean velocities ranging from $-6$ to 13 km s$^{-1}$. Several of them also overlap along the line of sight, as seen most clearly in substructures G, I, E, and H, consistent with the multiple spectral peaks observed in these areas. The approximate projected locations of the DR21 ridge and the W75N complex lie close to the interface between substructures G and I. In our FIVe decomposition, these two substructures have mean velocities of approximately $-0.2$ and $+8.2~\mathrm{km\,s^{-1}}$, respectively. These values are broadly consistent with the two well-known velocity systems in Cygnus~X North: the DR21 cloud at approximately $-3~\mathrm{km\,s^{-1}}$ and the W75N-related cloud at approximately $+9~\mathrm{km\,s^{-1}}$ \citep{2023NatAs...7..546S}.

\subsection{Turbulence in Substructures}

\begin{figure*}[htbp]
    \centering
    \includegraphics[width=1\textwidth]{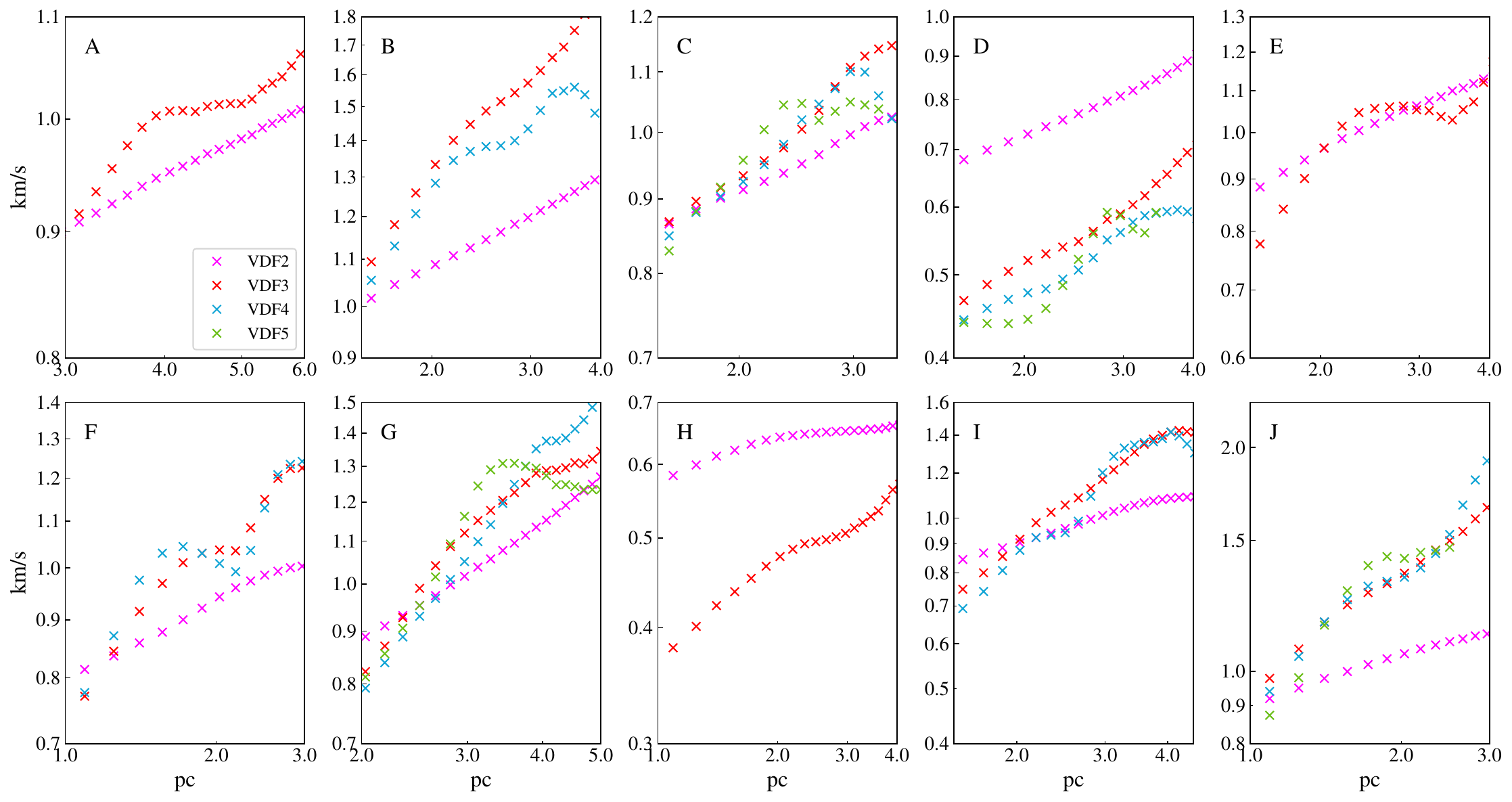}
    \caption{The two-point (magenta), three-point (red), four-point (blue), and five-point (green) VDFs around the plateau for each substructure. To keep the figure concise, once a lower-point VDF reveals a plateau, higher-point VDFs are omitted.}
    \label{fig:vdfs}
\end{figure*}

We compute the 2- to 5-point VDFs for the ten substructures following the same procedure as that applied to Region~I and II. Fig.~\ref{fig:vdfs} shows a zoomed-in view of the VDFs around the plateau region. All substructures exhibit a similar behavior: the VDFs increase from small scales, flatten into a plateau, and then rise again with fluctuations at larger separations. The presence of the plateau in all components indicates a characteristic turbulence correlation scale of $\sim 2$--5~pc. This scale is comparable to that derived for Cygnus~X as a whole, suggesting that the $l_{s}$ is broadly consistent across different regions and substructures.

A similar characteristic scale was also reported by \citet{2011A&A...529A...1S}, who applied the $\Delta$-variance method to the FCRAO \thirteenco($J=1-0$) integrated-intensity map of Cygnus~X. Adopting a distance of $\sim 1.7$~kpc, they found characteristic scales of $\sim 4$~pc and $\sim 40$~pc. They interpreted the $\sim 4$~pc scale as possibly related to small-scale filamentary structures or energy injection from H\,II regions, while also noting that it may partly arise from radiative-transfer effects when \thirteenco($J=1-0$) becomes optically thick in high-column-density regions. Although our VDF analysis probes the velocity-centroid structure rather than the intensity structure, the agreement in the characteristic few-parsec scale suggests that parsec-scale structures play an important role in both the spatial and kinematic organization of Cygnus~X.

From Fig.~\ref{fig:vdfs}, we find that plateaus are already visible in the 3-point VDFs of sub-A, E, and H. Sub-B, F, and I require the 4-point VDFs to reveal clear plateaus, whereas the plateaus of sub-C, D, G, and J become apparent only in the 5-point VDFs. We find no simple correspondence between the required point $n$ of VDF and the size of a substructure. The plateau amplitudes also vary substantially among the substructures, indicating differences in the rms amplitudes of their turbulent velocity fluctuations. These variations may reflect differences in the intrinsic physical properties and kinematic complexity of the clouds.

For instance, the plateau amplitudes of sub-I and G are significantly higher than those of sub-D and H. The approximate projected locations of the DR21 ridge and the W75N complex lie close to the interface between sub-I and G. The DR21/W75N environment hosts massive star-forming regions, filamentary accretion structures, and dynamically interacting atomic and molecular gas components extending over a broad velocity range \citep{2010A&A...520A..49S,2012A&A...543L...3H,2022ApJ...927..106C,2023NatAs...7..546S,2023ApJ...948L..17L}. These processes may contribute to stronger small-scale velocity fluctuations and, consequently, to the relatively high VDF plateau amplitudes of sub-I and G. By contrast, no comparably prominent dense structures are apparent in sub-D and H, consistent with their lower plateau amplitudes.

\subsection{Supersonic Turbulence in Different Scales or Structures}

We estimate the sonic Mach number ($\mathcal{M}_s$) from the multi-point VDFs as:
\begin{equation}
M_{s} = \frac{\sqrt{3} \left\langle \Delta V_{c}^{2} \right\rangle^{\frac{1}{2}}}{\sqrt{2} \left\langle \Delta V_{\mathrm{th}}^{2} \right\rangle^{\frac{1}{2}}}
\end{equation}
Here, $\langle \Delta V_c^2 \rangle^{1/2}$ corresponds to the VDFs. Since our goal is to compare the turbulent properties among different regions or substructures, we use the VDFs that exhibit a clear plateau as a proxy for $\langle \Delta V_c^2 \rangle^{1/2}$ in Eq.~(7). Besides, $\langle \Delta V_{\mathrm{th}}^2 \rangle^{1/2}$ represents the thermal velocity dispersion:
\begin{equation}
    \left\langle \Delta V_{\mathrm{th}}^{2} \right\rangle^{\frac{1}{2}} = \left( \frac{k_{\rm B} T}{\mu m_{H}} \right)^{\frac{1}{2}}
\end{equation}
where $k_{\rm B}$ is the Boltzmann constant, $T$ is the gas temperature, $\mu$ = 2.37 \citep{2008A&A...487..993K} is the mean molecular weight per free particle, and $m_{\rm H}$ is the atomic mass of hydrogen. 

\begin{figure*}[htbp]
    \centering
    \includegraphics[width=0.9\textwidth]{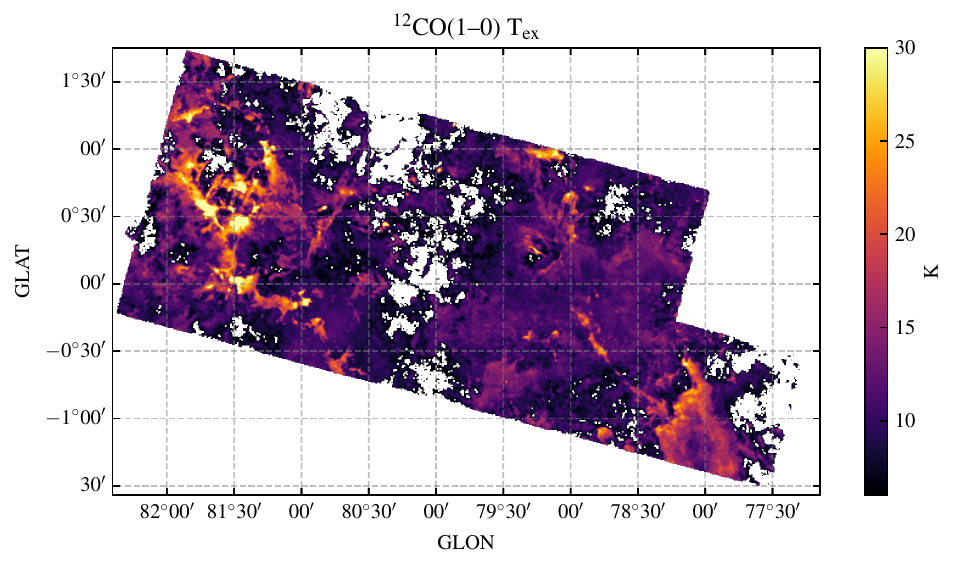}
    \caption{$T_{\rm ex}$ map of Cygnus~X derived from the \twelveco($J=1-0$) data. The $T_{\rm ex}$ is calculated from the peak temperature of the \twelveco($J=1-0$) spectrum within the velocity range from $-10$ to $20~\mathrm{km\,s^{-1}}$, assuming optically thick emission and a beam filling factor close to unity.}
    \label{fig:T}
\end{figure*}

Following \citet{2016A&A...587A..74S}, we estimate the gas excitation temperature ($T_{\rm ex}$) using the \twelveco($J=1-0$) line, which traces the most extended molecular gas component in Cygnus~X. Assuming that the \twelveco($J=1-0$) emission is optically thick and that the beam filling factor is close to unity, $T_{\rm ex}$ can be derived from the standard radiative-transfer equation \citep[e.g., Eq.~88 of][]{2015PASP..127..266M}. Substituting the \twelveco($J=1-0$) transition temperature $h\nu/k=5.53$~K and the cosmic microwave background temperature $T_{\rm bg}=2.73$~K gives
\begin{equation}
    T_{\rm ex}(^{12}{\rm CO}) =
    \frac{5.53}{\ln\left[1+\dfrac{5.53}{T_{\rm mb,peak}(^{12}{\rm CO})+0.818}\right]}
\end{equation}
Here, $T_{\rm mb,peak}(^{12}{\rm CO})$ is taken as the peak main-beam temperature of the \twelveco($J=1-0$) spectrum within the velocity range from $-10$ to $20~\mathrm{km\,s^{-1}}$. Fig.~\ref{fig:T} shows the resulting $T_{\rm ex}$ map of the main velocity component. The derived $T_{\rm ex}$ distribution is broadly consistent with the result presented by \citet{2016A&A...587A..74S} for Cygnus~X North (see their Fig.~B.1). Tab.~\ref{tab:1} lists the mean excitation temperatures of the different regions and velocity substructures measured from Fig.~\ref{fig:T}. The resulting temperature map shows that Region~I is systematically warmer than Region~II, with mean excitation temperatures of 16.1 and 13.9~K, respectively. Within Region~I, substructures E, G, and J have relatively high mean temperatures, reaching $\sim 20$~K, whereas sub-H has the lowest value of $\sim 12$~K. In Region~II, sub-B is noticeably warmer ($\sim18.3$~K) than the other three substructures. We use these \twelveco-derived $T_{\rm ex}$ as approximate gas temperatures in the Mach-number calculation.

\begin{deluxetable*}{rllllllllllll}
\tablewidth{0pt}
\tablecaption{The Mean Temperature of Different Regions and Structures \label{tab:1}}
\tablehead{
\colhead{Name} & \colhead{Region~I} & \colhead{Region~II} &
\colhead{A} & \colhead{B} & \colhead{C} & \colhead{D} &
\colhead{E} & \colhead{F} & \colhead{G} & \colhead{H} &
\colhead{I} & \colhead{J}
}
\startdata
$T_{\mathrm{mean}}$ (K) & 16.1 & 13.9 & 12.2 & 18.3 & 13.3 & 12.5 & 21.6 & 16.4 & 20.0 & 12.0 & 17.8 & 21.3 \\
\enddata
\end{deluxetable*}

\begin{figure*}[htbp]
    \centering
    \includegraphics[width=0.8\textwidth]{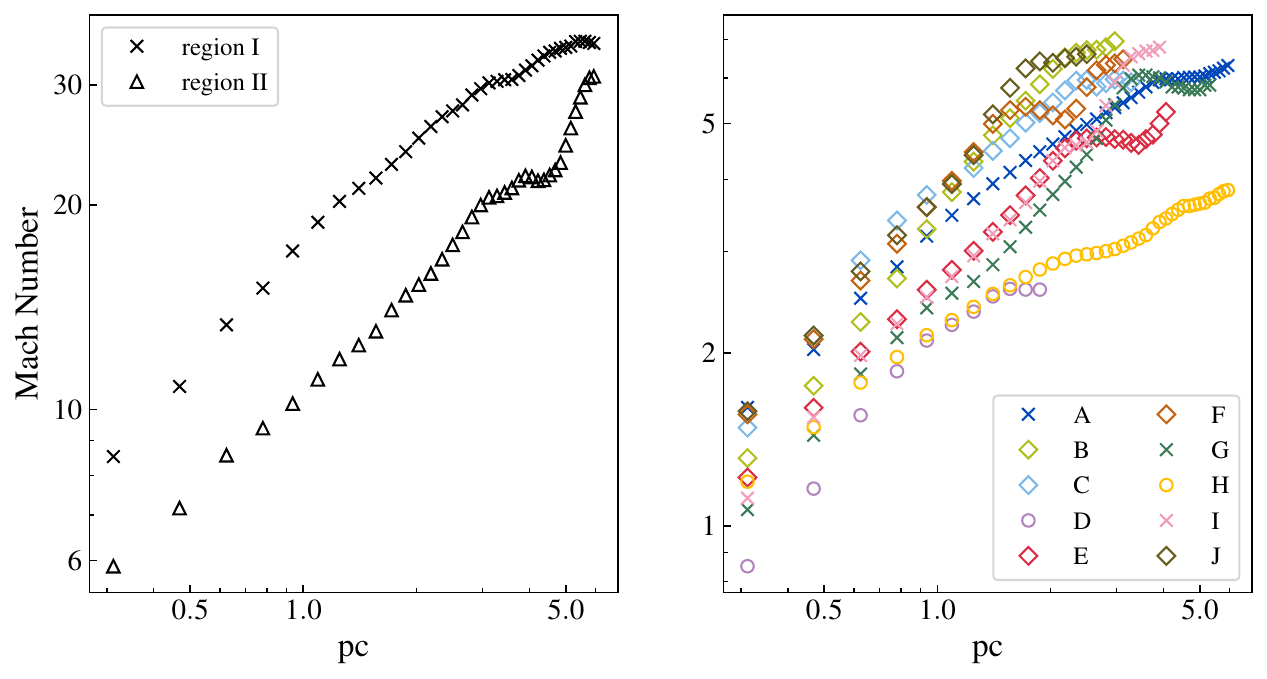}
    \caption{Sonic Mach number $\mathcal{M}_{s}$ as a function of separation scale. Left: Mach numbers in the two main regions of Cygnus~X, where crosses denote Region~I and triangles denote Region~II. Right: Mach numbers in the ten velocity substructures. Colors distinguish different substructures, while marker shapes indicate their morphologies: circles represent spherical-like clouds, diamonds represent filamentary-like clouds, and crosses indicate transitional shapes.}
    \label{fig:mach}
\end{figure*}

The left panel of Fig.~\ref{fig:mach} shows the calculated Mach number ($\mathcal{M}_s$) in the two main regions of Cygnus~X. Since the VDF$_{\rm 5pt}$ begins to fluctuate at $\sim 6$~pc, we restrict the analysis to scales below this limit, where a well-defined plateau is present. The plateau levels correspond to $\mathcal{M}_s \sim 30$ in Region~I and $\mathcal{M}_s \sim 20$ in Region~II, indicating highly supersonic turbulent motions. These results suggest that Cygnus~X is characterized by strongly supersonic turbulence, which can generate significant density fluctuations in the gas and is consistent with an active and dynamically evolving cloud.

The right panel of Fig.~\ref{fig:mach} demonstrates the Mach numbers from the VDFs of the ten velocity substructures. For each substructure, the range in scale is limited by the extent of the plateau. All substructures exhibit supersonic turbulence. In particular, substructures D and H show relatively low values of $\mathcal{M}_s \sim 2$--3, whereas the remaining substructures have higher values, typically $\mathcal{M}_s \sim 5$--7. A similar trend is seen in the comparison between the VDFs of Regions~I and II and those of the individual substructures (Figs.~\ref{fig:vdfl} and \ref{fig:vdfs}). However, the Mach numbers in the substructures are systematically lower than those measured in the larger regions. The physical origin of this difference is discussed in Section~\ref{sec:4.2}.

Notably, the two substructures with the lowest Mach numbers, sub-D and sub-H, both exhibit approximately spherical morphologies, suggesting that cloud shape may be related to the observed velocity dispersion and, consequently, the Mach number. In contrast, substructures with more complex geometries, such as sub-B and sub-J, tend to show significantly higher velocity dispersions. A similar trend was reported by \citet{2012MNRAS.425..720S} in their study of the dense interstellar medium in the Central Molecular Zone. One possible explanation is that more complex geometries introduce a broader range of velocity components along the line of sight, leading to larger measured dispersions. In addition, star formation and other dynamical processes may drive the morphological evolution of molecular clouds from initially more regular configurations toward filamentary or irregular structures \citep{2004RvMP...76..125M,2012ARA&A..50...29C}.

\section{Discussion} \label{sec:discussion}

\subsection{The Power Law of VDFs} \label{sec:4.1}

Previous studies have employed the linewidth--size relation to characterize turbulence in molecular clouds. In theory, isotropic and incompressible turbulence is expected to exhibit a scaling exponent of $1/3$, while compressible turbulence predicts a steeper exponent of $\sim 1/2$ \citep{2004ARA&A..42..275S,2007ARA&A..45..565M}. Observationally, reported exponents typically span a broad range from $\sim 0.2$ to $0.7$. However, linewidth measurements can be broadened by thermal motions and ordered kinematic components, which may bias the inferred scaling relation with respect to the intrinsic turbulent properties.

In this work, we adopt the VDFs as an alternative diagnostic. The 2-point VDF is closely related to the classical $\sigma$--$R$ relation: while the former characterizes velocity scaling within individual clouds and the latter compares different clouds, both probe the same underlying turbulent physics. In the right panels of Fig.~\ref{fig:vdfl}, we fit the 2- to 5-point VDFs with power laws for Regions~I and II, considering only spatial scales up to the turbulence correlation length $l_s$ to focus on the turbulent cascade with minimal influence from ordered motions. For 2-point VDF, we obtain power-law exponents of 0.45 and 0.41 for Regions~I and II, respectively. These values are consistent with previous observational results \citep{1983ApJ...270..105M,1987ApJ...319..730S,1995ApJ...446..665C,1998ApJ...504..223G,2004ApJ...615L..45H} and fall within the range expected for both incompressible and compressible turbulence.

As discussed above, however, 2-point VDF alone is not sufficient to remove ordered velocity components. As a result, even at scales smaller than $l_s$, the inferred scaling may still be affected by residual ordered motions. As the number of points in the VDF increases, its ability to suppress such components improves, making higher-point VDFs more suitable for probing intrinsic turbulence scaling. Consistent with this expectation, we find that the fitted power-law exponents increase systematically with VDF order in both regions. In Region~I, the exponent increases from 0.45 (2-point VDF) to 0.50, 0.51, and 0.51 for 3, 4, and 5-point VDF, respectively. In Region~II, the corresponding values rise from 0.41 to 0.48, 0.53, and 0.55.

\begin{figure*}[htbp]
    \centering
    \includegraphics[width=1\textwidth]{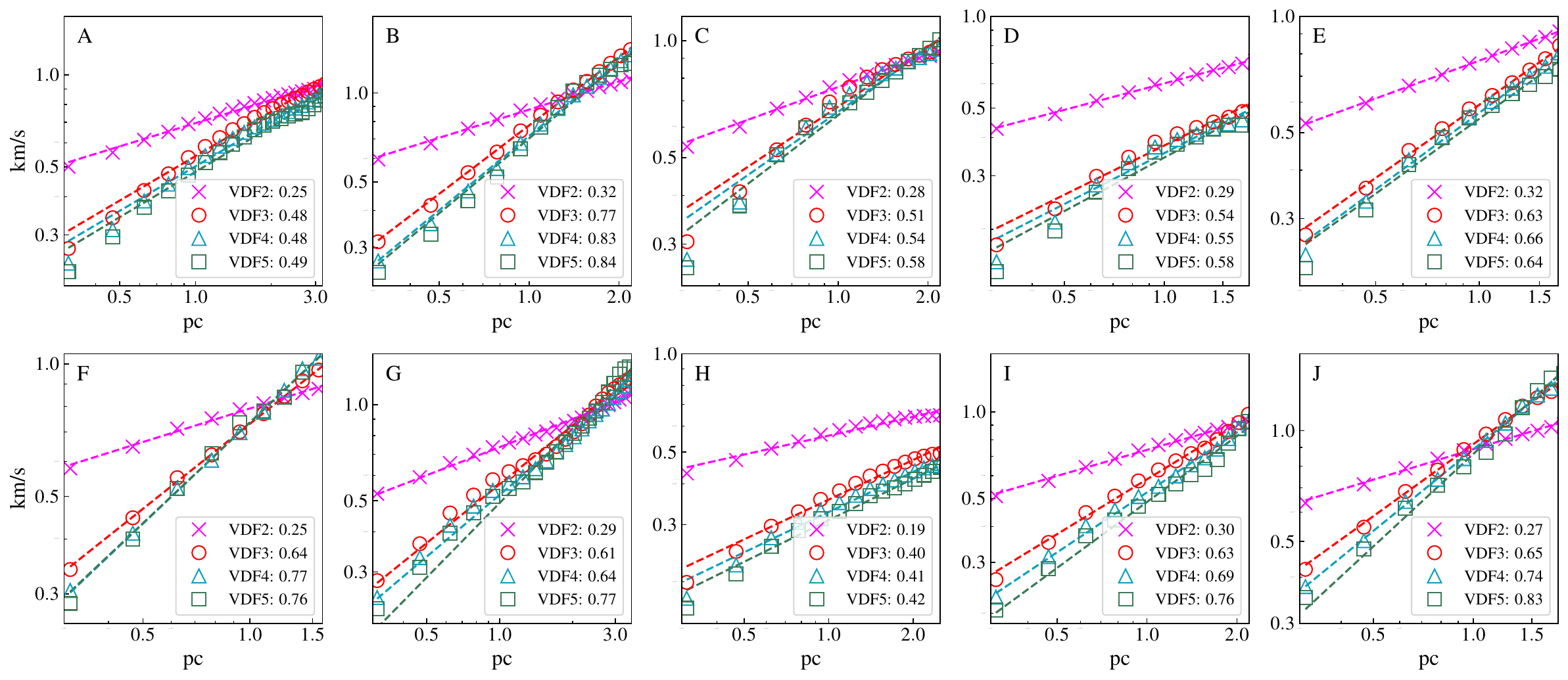}
    \caption{2-point VDFs (cross), 3-point VDFs (circle), 4-point VDFs (triangle), and 5-point VDFs (square) for each velocity component. The dashed line represents the result of fitting the multi-point VDFs, with the exponents listed in each panel.}
    \label{fig:powerlaw}
\end{figure*}

We further examine this behavior in individual substructures. Fig.~\ref{fig:powerlaw} presents the 2- to 5-point VDFs for the ten substructures, together with their power-law fits. In all cases, the power-law exponents increase from 2- to 5-point VDF, consistent with the trend observed in the larger regions. The exponents derived from 2-point VDF are relatively small, ranging from 0.19 to 0.32, and differ significantly from those measured for the entire Cygnus~X region. In contrast, the exponents obtained from higher-point VDFs are substantially larger; for example, the VDF$_{\rm 5pt}$ exponents in sub-B and sub-J exceed 0.8. The differences in scaling behavior between cloud complexes (Regions~I and II) and the individual clouds (10 substructures) are evident, which is discussed further in Section~\ref{sec:4.2}.

\subsection{The Correlation Scale and Velocity Dispersion Revealed by the VDFs} \label{sec:4.2}

As discussed above, the velocity field of a molecular cloud can be regarded as a superposition of large- and small-scale components, each of which may contain both ordered and random motions. Multi-point VDFs, especially the 4- and 5-point VDFs, are more effective at suppressing ordered velocity components on relatively small scales. As a result, when the separation approaches and slightly exceeds $l_s$, the VDF tends to flatten and form a plateau. This plateau can be interpreted as the dominant contribution from small-scale turbulent velocity fluctuations. At larger separations, the VDF rises again and may exhibit fluctuations, indicating that larger-scale velocity structures not fully removed by a finite-order VDF begin to dominate. 

Tab.~\ref{tab:plateau} lists the plateau scales and amplitudes of the VDFs for Regions~I and II and the 10 substructures, revealing two main results. First, their correlation lengths are not very different: the characteristic plateau scale is $\sim 3$--5~pc in Regions~I and II, and $\sim 2$--4~pc in the substructures, remaining within the same order of magnitude. Second, the plateau amplitudes, which correspond to the turbulence velocity dispersion and hence reflect the turbulent energy density, differ much more substantially. The plateau amplitudes are approximately $6~\mathrm{km\,s^{-1}}$ and $4~\mathrm{km\,s^{-1}}$ in Regions~I and II, respectively, whereas those of the substructures are typically only $\sim 0.5$--1.5~$\mathrm{km\,s^{-1}}$. Thus, the plateau levels in Regions~I and II are generally higher than those of the substructures by factors of roughly 4--8.

\begin{table}[t]
\centering
\caption{Approximate plateau scales and amplitudes of the VDFs for the two main regions and the 10 substructures. These values are rough estimates based on visual inspection of the plateau ranges in the VDFs.}
\label{tab:plateau}
\small
\begin{tabular}{lcc}
\hline
Name & Plateau scale (pc) & Amplitude ($\mathrm{km\,s^{-1}}$) \\
\hline
Region~I  & $\sim 3$--4 & $\sim 6.0$ \\
Region~II & $\sim 4$--5 & $\sim 4.0$ \\
sub-A     & $\sim 3$--4 & $\sim 1.0$ \\
sub-B     & $\sim 2$--3 & $\sim 1.4$ \\
sub-C     & $\sim 2$--3 & $\sim 1.0$ \\
sub-D     & $\sim 3$    & $\sim 0.6$ \\
sub-E     & $\sim 2$--3 & $\sim 1.1$ \\
sub-F     & $\sim 2$    & $\sim 1.0$ \\
sub-G     & $\sim 3$--4 & $\sim 1.3$ \\
sub-H     & $\sim 2$--3 & $\sim 0.5$ \\
sub-I     & $\sim 3$--4 & $\sim 1.4$ \\
sub-J     & $\sim 2$    & $\sim 1.5$ \\
\hline
\end{tabular}
\end{table}

If the differences between Regions~I and II and the 10 substructures were interpreted simply as turbulence measured at different scales of the cascade, then the corresponding turbulence velocity dispersion $\sigma_s$ and correlation scale $l_s$ would be expected to follow the usual turbulence scaling, roughly $\sigma_s \propto l_s^{1/3}$--$l_s^{1/2}$. In that case, the relatively small differences in $l_s$ would not naturally lead to the much larger differences observed in the plateau amplitudes. This suggests that the contrast between the cloud complexes and the substructures cannot be explained solely in terms of a scale-dependent change in the intrinsic turbulence.

A more plausible interpretation is that, in real observational data, multi-point VDFs may be more robust in revealing the turbulence correlation scale than in measuring the turbulence velocity dispersion. For Regions~I and II and the 10 substructures, the plateau scales differ only slightly, suggesting that the VDF can steadily identify the turbulence correlation length. By contrast, the plateau amplitudes, which correspond to $\sigma_s$, may be more sensitive to residual large-scale motions that are not fully removed by the finite-order VDF. In this sense, the similar $l_s$ values may more reliably indicate a common turbulence correlation scale in the present data, whereas the much larger $\sigma_s$ in Region~I and II than the substructures may include additional contributions from incompletely removed large-scale velocity structures rather than purely intrinsic differences in turbulent energy density. Clarifying this issue will require observations with both wider spatial coverage and a larger dynamical range, so that the VDF behavior can be better constrained both near the plateau and at separations larger than $l_s$.

\section{Conclusion} \label{sec:conclusion}

Turbulence plays a critical role in regulating star formation. Traditional methods for investigating turbulence properties, such as the linewidth--size relation, face challenges in massive molecular clouds due to thermal broadening and rotational effects. We applied multi-point VDFs to the Nobeyama \thirteenco\ $(J=1-0)$ data of Cygnus~X to investigate the turbulent properties of this massive molecular cloud. Our main conclusions are as follows:

\begin{itemize}
    \item Multi-point VDFs provide an effective way to suppress ordered velocity components and isolate the small-scale turbulent contribution. In Cygnus~X, clear plateau features are mainly identified in the 4-point and 5-point VDFs.

    \item The $l_{s}$ are $\sim 3$--4~pc in Region~I and $\sim 4$--5~pc in Region~II, with plateau amplitudes of $\sim 6$ and $\sim 4~\mathrm{km\,s^{-1}}$, respectively. These correspond to plateau-based sonic Mach numbers of $\mathcal{M}_s \sim 30$ and $\sim 20$.

    \item Using the FIVe algorithm, we identify 10 velocity substructures in Cygnus~X. Their plateau scales are $\sim 2$--4~pc, comparable to those of the two main regions, but their plateau amplitudes are much lower, typically $\sim 0.5$--1.5~$\mathrm{km\,s^{-1}}$, with Mach numbers of $\mathcal{M}_s \sim 2$--7.

    \item The power-law exponents increase systematically from the 2-point to the 5-point VDFs in both the cloud complexes Region~I and Region~II and the 10 substructures, consistent with progressively better removal of ordered velocity components.

    \item The similar correlation lengths but markedly different plateau amplitudes between cloud complexes and their substructures suggest that, in observational data, multi-point VDFs provide a more robust constraint on the turbulent correlation length than measurements of turbulent velocity dispersions.
\end{itemize}

\begin{acknowledgments}

This work is supported by the National Natural Science Foundation of China (NSFC) grant No. 12425304 and the National Key R\&D Program of China with Nos. 2023YFA1608204 and 2022YFA1603103. We are grateful to the Early Research Training Program of the School of Astronomy and Space Science, Nanjing University, for their strong support of this study. Our thanks also go to the Nobeyama Radio Observatory (NRO), with the Nobeyama 45m radio telescope operated by NRO, a branch of the National Astronomical Observatory of Japan. Data analysis was performed on the open-use data analysis computer system at the Astronomy Data Center (ADC), National Astronomical Observatory of Japan. J.L. was partially supported by Grant-in-Aid for Scientific Research (KAKENHI Number JP23H01221 and JP25K17445) of the Japan Society for the Promotion of Science (JSPS). M.Z. acknowledges support from the National Natural Science Foundation of China (NSFC) grant No. 12503029. This research made use of \texttt{astropy}, a community-developed core Python package for Astronomy \citep{2013A&A...558A..33A}, as well as \texttt{NumPy} and \texttt{SciPy} \citep{2011CSE....13b..22V}, and \texttt{Matplotlib} \citep{2007CSE.....9...90H}.

\end{acknowledgments}

\bibliography{sample701}{}
\bibliographystyle{aasjournalv7}

\appendix
\section{Effects of Angular and Velocity Resolution}
\label{app:resolution}

\begin{figure*}[htbp]
\centering
\includegraphics[width=\textwidth]{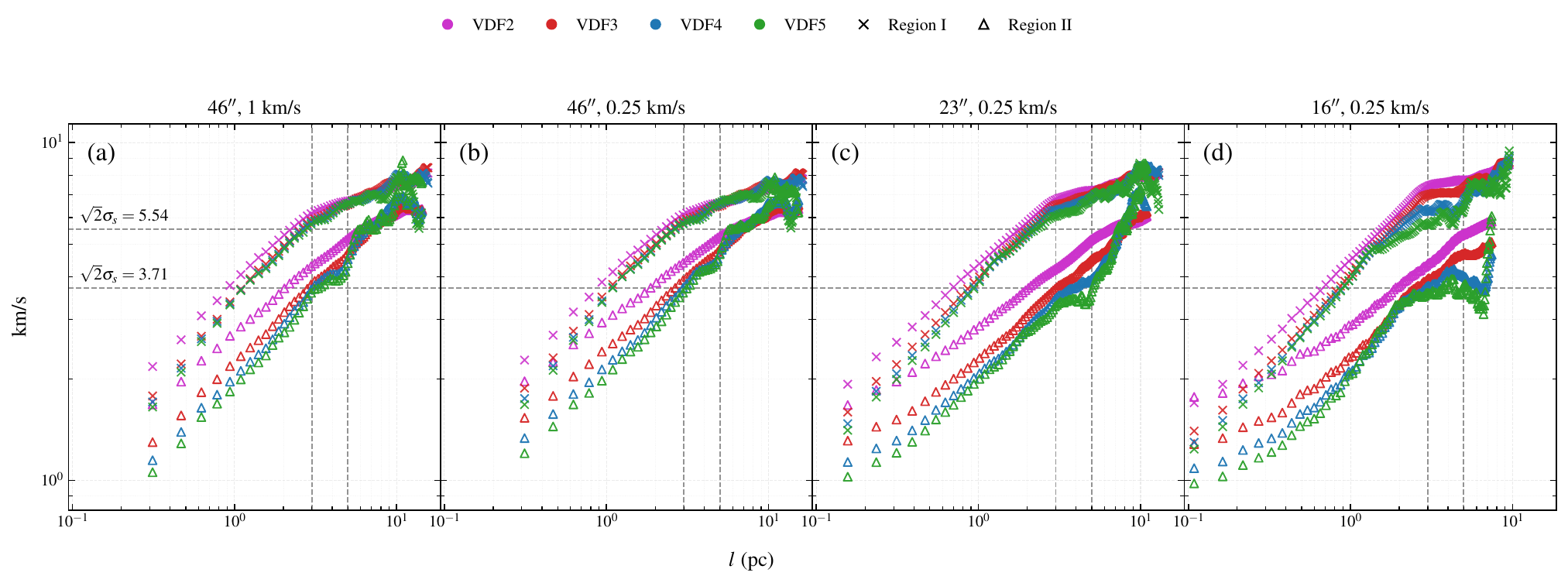}
\caption{Comparison of the VDFs derived from data products with different effective angular resolutions and velocity-channel widths. The symbols follow those used in Fig.~\ref{fig:vdfl}, and the title of each panel indicates the resolution of the corresponding data product. Panel~(a) shows the data adopted in the main analysis, while panels~(b)--(d) show the results obtained with a velocity-channel width of $0.25~\mathrm{km\,s^{-1}}$ and different angular resolutions. The horizontal dashed lines indicate $\sqrt{2}\sigma_s$ for Regions~I and II, while the vertical dashed lines mark spatial scales of 3 and 5~pc.}
\label{fig:9}
\end{figure*}

To assess the effects of angular and velocity resolution on the derived VDFs, we repeat the analysis for Regions~I and II using the Nobeyama 45m Cygnus~X CO survey products with a velocity-channel width of $0.25~\mathrm{km\,s^{-1}}$ and effective angular resolutions of $16^{\prime\prime}$, $23^{\prime\prime}$, and $46^{\prime\prime}$. These results are compared with those from the data adopted in the main analysis, which have an effective angular resolution of $46^{\prime\prime}$ and a velocity-channel width of $1.0~\mathrm{km\,s^{-1}}$. The resulting VDFs are shown in Fig.~\ref{fig:9}.

Overall, the VDFs derived from the different data products all show plateau features at scales of approximately 3--5~pc. By contrast, the plateau amplitudes and detailed shapes of the VDFs show modest differences. In particular, the higher-angular-resolution data exhibit stronger fluctuations at large separations. This may be because a finer effective angular resolution preserves more high-spatial-frequency structure, while finite and irregular spatial sampling further enhances the variations as the number and spatial distribution of valid multi-point configurations change with separation.

These comparisons indicate that the turbulence correlation scale inferred from the VDFs is not strongly affected by the adopted angular or velocity resolution, although the plateau amplitude and detailed profile shape are more sensitive to the data product. For our primary goal of using the VDFs to characterize the turbulence properties of Cygnus~X, the data with an effective angular resolution of $46^{\prime\prime}$ and a velocity-channel width of $1.0~\mathrm{km\,s^{-1}}$ are therefore sufficient. We adopt this data product in the main analysis because of the higher S/N and clearer and more stable VDF profiles.

\end{document}